\documentclass{iau}
\usepackage{graphicx}
\usepackage{amssymb}
\title[$C^3$: a Command-line Catalogue Cross-matching tool for modern astrophysical data] 
{C$^3$: A Command-line Catalogue Cross-matching tool for modern astrophysical survey data}

\author[Giuseppe Riccio et al.]   
{Giuseppe Riccio$^{1}$,  Massimo Brescia$^{1}$,   Stefano Cavuoti$^{1}$, Amata Mercurio$^{1}$,
Anna Maria Di Giorgio$^{2}$ \and  Sergio Molinari$^{2}$}

\affiliation{$^1$ INAF - Astronomical Observatory of Capodimonte, via Moiariello 16, I-80131 Napoli, Italy\\[\affilskip]
$^2$ INAF - Istituto di Astrofisica e Planetologia Spaziali, Via Fosso del Cavaliere 100, I-00133 Roma, Italy}

\pubyear{2017}
\volume{325}  
\jname{Astroinformatics 2016}
\editors{M. Brescia, S. Cavuoti, G.S. Djorgovski, \\ E. Feigelson \& G. Longo, eds.}
\begin{document}

\maketitle

\begin{abstract}
In the current data-driven science era, it is needed that data analysis techniques has to quickly evolve to face with data whose dimensions has increased up to the Petabyte scale. In particular, being modern astrophysics based on multi-wavelength data organized into large catalogues, it is crucial that the astronomical catalog cross-matching methods, strongly dependant from the catalogues size, must ensure efficiency, reliability and scalability. Furthermore, multi-band data are archived and reduced in different ways, so that the resulting catalogues may differ each other in formats, resolution, data structure, etc, thus requiring the highest generality of cross-matching features. 
We present $C^{3}$ (Command-line Catalogue Cross-match), a multi-platform application designed to efficiently cross-match massive catalogues from modern surveys. Conceived as a stand-alone command-line process or a module within generic data reduction/analysis pipeline, it provides the maximum flexibility, in terms of portability, configuration, coordinates and cross-matching types, ensuring high performance capabilities by using a multi-core parallel processing paradigm and a sky partitioning algorithm.
\end{abstract}
%
%
\firstsection 
\section{Introduction}
Today we are in an era, the so-called \textit{data-driven} era, in which the size of data has reached dimensions so huge that often it is  humanly impossible to handle them in an efficient and comprehensible way. Dealing with petabytes of data will be the standard in a not far future and this will require that data analysis techniques and facilities must quickly evolve to face such amount of information.

In particular, in astrophysics, the data volumes from the ongoing and next generation multi-band and multi-epoch surveys are expected to be so huge and diversified that analyzing, cross-correlating, visualizing and extracting knowledge from such data will represent a not trivial challenge for astronomers and computer engineers. The massive multi-band and multi-epoch information, foreseen to be available from the on-going and future surveys, will require efficient techniques and software solutions to be directly integrated into the reduction pipelines, making possible to cross-correlate in real time a large variety of parameters for billions of sky objects.

One of main core steps of any standard modern pipeline of astronomical data reduction/analysis is the cross-matching technique, consisting in associating and comparing sources belonging to different catalogues. The cross-matching among modern astronomical catalogues is a challenge for two main reasons: first, it is particularly sensible to the growing of the datasets size, if performed in a naive way; second, multi-band data, even if referred to a same sky region, are archived and reduced in different ways, so that the resulting catalogues may differ each other in formats, resolution, data structure, etc, thus requiring the highest generality of cross-matching features.

In this work we describe the features of C$^{3}$, \textit{Command-line Catalogue Cross-match}, (\cite{riccio2016})\footnote{The C$^{3}$ tool and the user guide are available at the page: http://dame.dsf.unina.it/c3.html.}, a tool to perform efficient cross-matching among heterogeneous catalogues from modern astronomical surveys. It can be easily integrated into an automatic data analysis pipeline and its high performance capabilities are ensured by the use of a multi-core parallel processing paradigm and a sky partitioning algorithm. Furthermore, this tool has been tailored on the specific user needs, giving the maximum flexibility to the end-user in terms of portability, parameter configuration, catalogue formats, angular resolution, region shapes, coordinate units and cross-matching types.

\section{C$^3$ design and architecture}
C$^{3}$, a command-line Python tool designed and developed to cross-match modern astrophysical catalogues, can be run as a stand-alone process or integrated within any generic data reduction/analysis pipeline. In order to ensure high performance in terms of computational time, it is based on a sky partitioning algorithm and on the multi-core parallel processing paradigm. Moreover, C$^3$ is aimed to meet the needs of a community as wide as possible by providing the maximum flexibility to the end-user, in terms of catalogue formats and parameters, coordinates systems and cross-matching functions; finally, it is very simple to configure through filling of a configuration file.

Summarizing, the main features of the C$^3$ tool are:

\begin{itemize}
 \item \textit{Command line}: it can be used as stand-alone process or integrated within complex data analysys pipelines;
 \item \textit{Python compatibility}: up to the version 3.4;
 \item \textit{Standard libraries and facilities}: it makes use of the standard astronomical tool STILTS (\cite[Taylor 2006]{taylor2006}) and two common Python libraries, \textit{NumPy} (\cite[Van der Walt et al. 2011]{vanderwalt2011}), and \textit{PyFITS}\footnote{PyFITS is a product of the Space Telescope Science Institute, which is operated by AURA for NASA.};
 \item \textit{Multi-platform}: C$^{3}$ is compatible with Ubuntu Linux (from rel.14.04), Windows 7/10, Mac OS and Fedora;
 \item \textit{Multi-process}: a multi-core parallel processing paradigm is used to improve performance in terms of computational time;
 \item \textit{Sky partitioning}: a sky partitioning algorithm is used to reduce computational time;
 \item \textit{Astronomical I/O formats compliant}: the tool works with the most common I/O formats: FITS, ASCII, CSV, VOTable;
 \item \textit{Standard astronomical coordinate systems compliant}: C$^3$ works with equatorial (icrs, fk4, fk5) and galactic coordinate systems, expressed in degrees, radians or sexagesimal;
 \item \textit{User-friendliness}: only a simple configuration file is required to configure and use it.

\end{itemize}

\subsection{Cross-matching use cases}

C$^{3}$ provides three different use cases, according to the most common cross-matching criteria used by the astronomical community:

\begin{enumerate}
 \item \textit{Sky}: two objects from two different catalogues match if they fall within the same elliptical or rectangular sky area defined by the catalogue parameters. It is defined \textit{positional crossmatch};
 \item \textit{Exact Value}: two objects are matched if they have the same value for a pair of columns (one for each catalogue) defined by the user;
  \item \textit{Row-by-Row}: the cross-match is performed on a same row-ID of the two catalogues.
\end{enumerate}

\begin{figure}
\centering
  \includegraphics[width=6cm]{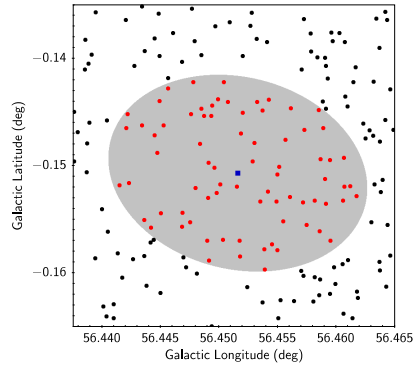}\\

\caption{An elliptical cross-match between two catalogues: the grey ellipse is the elliptical matching region centred in the coordinates of an object of first catalogue (squared dot in the center of the ellipse); The objects (red or light grey dots), belonging to the second catalogue and falling into such region are matching with the central object.}\label{fig:crossmatch}
\end{figure}

The idea at the basis of the C$^3$ positional cross-matching method for the \textit{Sky} use case is the same of the Q-FULLTREE approach, an our tool designed and developed in the FP7 ViaLactea project framework, introduced in \cite{becciani2015} and \cite{sciacca2016}: the coordinates of each object of the first catalogue identify the centre of an elliptical (circular as special case) or rectangular region having fixed dimensions or defined by catalogue parameters (for example, the two FWHMs can be used as semi-axis of the ellipse or as width and height of the rectangle). By calculating the distance of each object of the second catalogue from the centre of region, it is possible to evaluate if it falls into such region. In the affermative case, the two objects are matching (Fig.~\ref{fig:crossmatch}) .

\begin{figure*}
\centering
\includegraphics[width=10cm]{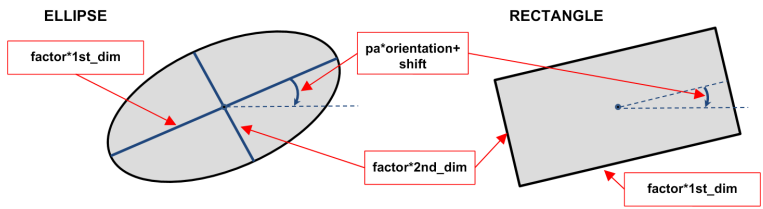}

\caption{Configuration of C$^{3}$ matching area: it can be elliptical (circular as special case) or rectangular; its size, representing  the ellipse axes or width and height of the rectangle, are defined by a fixed value or by specific columns of the catalogues multiplied by a user defined \textit{parametric factor}; the position angle is characterized by a value and two additional parameters opportunely set the zero-point and the direction of rotation, namely, \textit{shift} and \textit{orientation}}\label{fig:pa}
\end{figure*}

The parameters characterizing the matching region can be defined by the user in the configuration file. In particular, it is possible to set: the shape of the matching area, which can be elliptical or rectangular (circular is a special elliptical case); the dimensions of the area, which can be defined by a fixed value or extracted by specific columns of the catalogues multiplied by a user-defined factor; its orientation, characterized by a positional angle (defined by a fixed value or by catalogue parameters) and two additional parameters, to opportunely set the zero-point and the direction of rotation (Fig.~\ref{fig:pa}).

The user has also to specify which matches must be included in the output file (\textit{all} the matches or only the \textit{best} pairs, in the sense of closest objects), according to the logical rule they have to satisfy:

\begin{description}
 \item[$1$ and $2$], only rows having an entry in both input catalogues;
 \item[$1$ or $2$], all rows, matched and unmatched, from both input catalogues;
 \item[All from $1$ (All from $2$)], all matched rows from catalogue $1$ (or $2$), together with the unmatched rows from catalogue $1$ (or $2$);
 \item[$1$ not $2$ ($2$ not $1$)], all the rows of catalogue $1$ (or $2$) without matches in the catalogue $2$ (or $1$);
 \item[$1$ xor $2$], the ``exclusive or'' of the match - i.e. only rows from the catalogue $1$ not having matches in the catalogue $2$ and viceversa.
\end{description}

\subsection{Performance Boosters}
In order to have high performance in terms of computational time, C$^3$ makes use of two different methods: \textit{i)} the application of multi-core parallel processing paradigm, through the definition of the number of parallel processes to run; \textit{ii)} a sky partitioning algorithm to reduce the number of comparisons between the sources of the two catalogues. 

\begin{figure}
\centering
  \includegraphics[width=8cm]{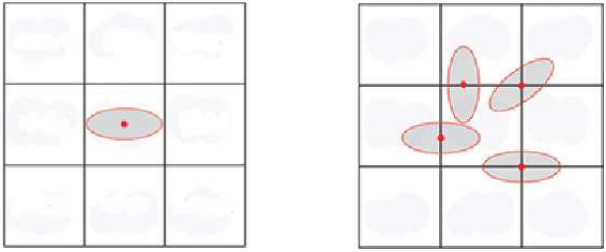}\\

\caption{The $C^{3}$ sky partitioning method. The sky is divided in cells whose dimensions are determined by the maximum value assumed by the main dimension of the matching area or by the \textit{minimum partition cell size} parameter (panel a). Each object of the second catalogue is assigned to a cell: a match between a source and the ellipse associated to the first catalogue object can take place only in the nine cells surrounding it (panel b).}\label{fig:partitioning}
\end{figure}

In order to apply the two methods, C$^{3}$ performs a series of preparatory manipulations on input data: the first input catalogue is splitted into a number of subsamples such that each of them is assigned to a concurrent process; then, the sky is divided into \textit{cells}, whose dimensions are defined by the maximum dimension that the matching regions can assume, namely the \textit{minimum cell size}, (Fig.~\ref{fig:partitioning}a). Each object is then assigned to a cell according to its coordinates. This choice of the unit cell dimensions has the result that a match between two objects can happen only if a source of the second catalogue lies in the nine cells surrounding the object of the first one (also known as Moore's neighborhood, \cite{gray2003}, see also Fig.~\ref{fig:partitioning}b), so avoiding the so-called \textit{block-edge problem} (\cite[Du et al. 2015]{du2014}).


In order to have the best performance for the specific computer used in the experiments, the user can manage both the number of concurrent processes and the dimension of the unit cell by setting their value in the configuration file.

\section{Testing}

In order to evaluate the C$^3$ performance in terms of computational time efficiency, we performed an intensive test campaign on real data. We used two catalogues extracted from the UKIDSS GPS public data, (\cite[Lucas et al. 2008]{lucas2008}), and the GLIMPSE Spitzer Data, (\cite{benjamin2003}, \cite{churchwell2009}), in the range of galactic coordinates $l\in[40,50]$, $b\in[-1,1]$. The tests have been performed by varying the dimensions of subsets and combining each subsample of the first catalogue with all the subsamples of the second one. In particular, datasets with, respectively, $1000$, $10,000$, $100,000$, $1,000,000$ and $10,000,000$ objects have been created from the UKIDSS catalogue, while, from the GLIMPSE catalogue, datasets with $1000$, $10,000$, $100,000$ and $1,000,000$ rows have been extracted.

The computer used for the tests is equipped with an Intel(R) Core(TM) $i5-4460$, with one $3.20GHz$, $4-core$ CPU, $32$ GB of RAM and hosting Ubuntu Linux $14.04$ as operative system on a standard Hard Disk Drive.

\begin{figure}
\centering
  \includegraphics[width=13.5cm]{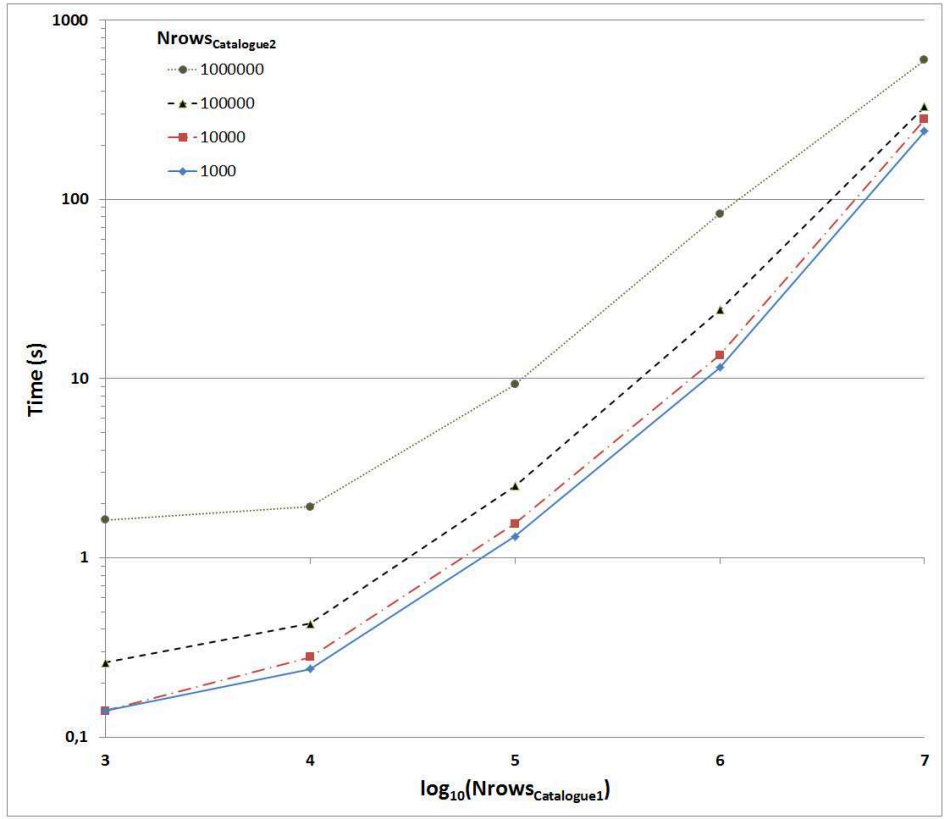}\\

\caption{Computational time trends of cross-matching phase as function of the number of rows of the first input catalogue for four different dimensions of the second catalogue: $1000$ rows (diamonds), $10,000$ rows (squares), $100,000$ rows (triangles), $1,000,000$ rows (circles).}\label{fig:c3test}
\end{figure}

The results of the tests are reported in Fig.~\ref{fig:c3test}. Each line in the figure represent the trend of computational time of the C$^{3}$ cross-matching phase as function of the incremental number of objects in the first catalogue for a fixed number of rows of the second catalogue (from $1000$ to $1,000,000$ rows as previously described). The results are in a perfect agreement with other publicly available cross-matching tools, as shown in \cite[Riccio et al. (2016)]{riccio2016}.

\section{Conclusions}

In this paper we have described a new scalable tool, named C$^3$, designed and developed to perform the cross-matching between astronomical datasets. It is a multi-platform command-line Python script, executable as a stand-alone software or as a module in a generic data reduction/analysis pipeline, whose high performance capabilities are guaranteed by the use of the multi-core parallel processing paradigm and an indexing function to partionate the sky.
The tool provides a number of use cases and parameter choice (I/O formats, coordinates systems, shape and dimensions of matching area and cross-matching type) in order to meet the needs of a scientific community as wide as possible.
Finally, C$^3$ requires only a configuration file to run, making it easy to use.

An intensive test campaign on real data has demonstrated its reliability and good performance capabilities, in particular when input catalogues increase their dimensions.

The C$^{3}$ tool, (\cite{riccio2016}), and the user guide are available at the webpage http://dame.dsf.unina.it/c3.html.


\begin{thebibliography}{99}

\bibitem[Becciani et al. (2015)]{becciani2015} 
{U. Becciani, et al.} 2015
{\em Advanced Environment for Knowledge Discovery in the VIALACTEA Project}, Proceedings of ADASS XXV conference, October 2015, Sidney, Australia, in press. eprint arXiv:1511.08619

\bibitem[Benjamin et al. 2003]{benjamin2003} 
{R.~A. Benjamin, et al.} 2003
{\em The Publications of the Astronomical Society of the Pacific}, 115(810), 953-964

\bibitem[Churchwell et al. 2009]{churchwell2009}
{E. Churchwell, et al.} 2009
{\em Publications of the Astronomical Society of Pacific}, 121(877), 213-230

\bibitem[Du et al. (2014)]{du2014} 
{P. Du, et al.}, 2014
{\em Science China Physics, Mechanics and Astronomy}, 57(3), 577-583

\bibitem[Gray 2003]{gray2003} 
{L. Gray} 2003
{\em Not. Amer. Math. Soc.}, 50, 200-211

\bibitem[Lucas et al. (2008)]{lucas2008} 
{P.~W Lucas, et al.} 2008
{\em Monthly Notices of the Royal Astronomical Society}, 391(1), 136-163

\bibitem[Riccio et al. 2016]{riccio2016}
{G. Riccio, et al}{2016}
accepted for publication in {\em The Publications of the Astronomical Society of the Pacific}, eprint arXiv:1611.04431

\bibitem[Sciacca et al. (2016)]{sciacca2016} 
{E. Sciacca, et al.} 2016
{\em Milky Way analysis through a Science Gateway: Workflows and Resource Monitoring}, Proceedings of 8th International Workshop on Science Gateways, June 2016, Rome, Italy, submitted.

\bibitem[Taylor (2006)]{taylor2006}
{M.~B. Taylor} 2006
in Astronomical Data Analysis Software and Systems XV, ed. C. Gabriel, C. Arviset, D. Ponz, \& S. Enrique, {\em Astronomical Society of the Pacific Conference Series}, 351, 666.

\bibitem[Van Der Walt et al. (2011)]{vanderwalt2011}
{S. Van Der Walt, et al.} 2011
{\em Computing in Science and Engineering}, 13, 22







\end{thebibliography}
\end{document}